\newcommand{\CLASSINPUTtoptextmargin}{0.78in}
\newcommand{\CLASSINPUTbottomtextmargin}{1.06in}
\documentclass[conference,letterpaper]{IEEEtran}
\IEEEoverridecommandlockouts
\usepackage{amsmath,amssymb,amsfonts}
\usepackage{graphicx}
\usepackage{booktabs}
\usepackage{array}
\usepackage{xcolor}
\usepackage{hyperref}
\usepackage{microtype}
\usepackage{textcomp}
\usepackage[sorting=none]{biblatex}

\title{RACE-AIMC: Selective Inference for Heterogeneous Analog In-Memory Accelerators at the Edge}

\author{\IEEEauthorblockN{Osama Yousuf}
\IEEEauthorblockA{WD Research \\
San Jose, CA, USA \\
osama.yousuf@wdc.com}
\and
\IEEEauthorblockN{Martin Lueker-Boden}
\IEEEauthorblockA{WD Research \\
San Jose, CA, USA \\
martin.lueker-boden@wdc.com}
}

\begin{document}
\maketitle

\begin{abstract}
Analog in-memory computing (AIMC) speeds up neural-network inference by
doing the arithmetic directly inside a memory array, instead of shuttling
weights back and forth between memory and a processor. This saves energy,
but the physical devices that store the weights are imperfect: programming
errors, electrical noise, limited-resolution converters, and outright
broken cells all distort the computation, and every physical chip is
distorted in its own way. A designer with several such chips available
faces an uncomfortable choice: run all of them and combine the answers
(safe, but wasteful of energy), or trust a single chip blindly (cheap, but
with no guarantee on how often it is wrong). This paper introduces \textbf{RACE-AIMC} (\emph{Risk-Aware Certified
Ensemble for AIMC}), a framework that resolves this choice with statistics
rather than guesswork. Offline, RACE-AIMC studies a pool of physical
accelerators, picks the single best one for a given energy budget, and
computes a mathematically exact upper bound on how often that accelerator
will be wrong \emph{when it chooses to answer}. Online, only that one
accelerator is switched on; a lightweight check decides whether to accept
its answer or defer to a fallback. In our simulations using a noisy weight
mapping and multiple independent test runs, every certified bound stayed under
a 10\% error target (mean bound 7.83\% ± 0.89\%, with 70.88\% ±
0.98\% of inputs answered directly). The resulting system matches the
accuracy of a clean digital baseline while cutting
modeled energy use by 69.02\% relative to always running every accelerator
in the pool.
\end{abstract}


\begin{figure*}[htbp]
\centering
\includegraphics[width=0.8\textwidth]{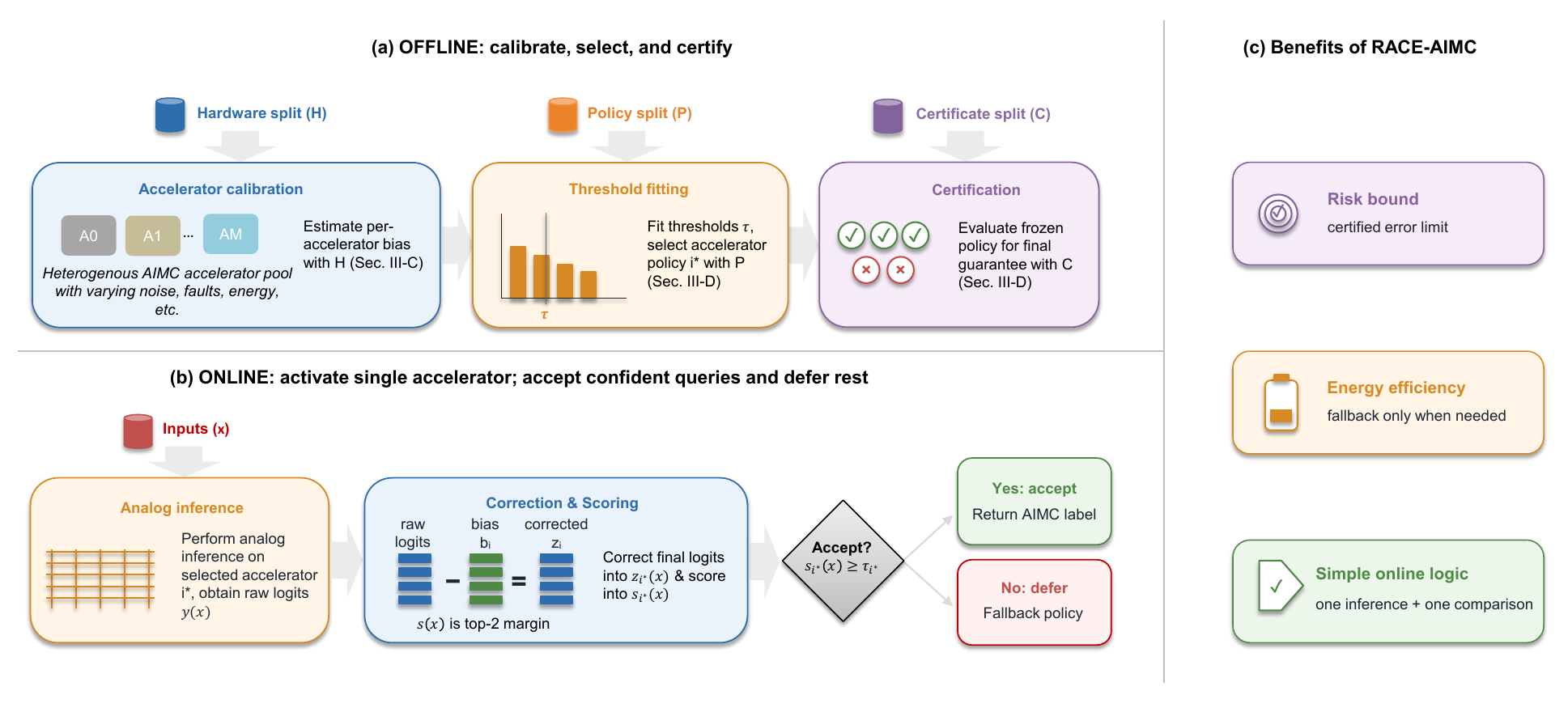}
\caption{RACE-AIMC overview. (a) Offline, disjoint hardware, policy, and
certificate splits characterize a heterogeneous pool of AIMC
accelerators, select and freeze one accelerator and its confidence threshold,
and independently certify accepted error. (b) Online, a query activates only the
selected accelerator; a small controller accepts its high-confidence label or
defers to a fallback or abstention path. (c) Benefits of our approach. }
\label{fig:overview}
\end{figure*}

\section{Introduction}
Ordinary computers keep memory (where numbers are stored) and the
processor (where arithmetic happens) in physically separate places.
Every multiplication requires shuttling a number from memory to the
processor and back, which costs both time and energy. Analog
in-memory computing (AIMC) avoids most of that shuttling: neural-network
weights are stored as physical \emph{conductances} inside a grid of
memory devices (a ``crossbar array''), and a matrix-vector multiplication
-- the core operation in a neural network layer -- is computed directly by
applying voltages to the array and reading off currents, using basic
circuit physics (Ohm's and Kirchhoff's laws) instead of digital
multiply-and-add circuits \cite{sebastian2020memory, le202364}.

The catch is that physical devices are not perfect. Writing a target
conductance value onto a device is imprecise (\emph{programming error});
reading the device back introduces \emph{noise}; converting between the
analog and digital worlds adds \emph{quantization error}; and a small
fraction of devices may be stuck at the wrong value entirely (\emph{stuck
faults}) \cite{rasch2023hardware, wang2022device, yousuf2025layer}. All of this means a neural network mapped
onto an analog chip does not compute quite the same thing as the original
digital network -- and because every physical chip has its own random
manufacturing imperfections, \emph{different chips make different
mistakes}.

If you have several such imperfect chips available (a \emph{heterogeneous
pool}), an obvious safety net is to run every input through all of them
and combine the results, e.g.\ by averaging. This is the standard
engineering answer to unreliable hardware, but on an energy-constrained
edge device it is wasteful: you pay the full energy cost on \emph{every
single input}, even the easy ones a single chip would have gotten right
anyway. The opposite extreme -- picking one chip and trusting it blindly
-- is cheap, but gives no way to guarantee an application-level error
budget (e.g., ``this system must be wrong on at most 10\% of the answers
it commits to'').

A middle path, borrowed from the idea of {selective classification}
\cite{geifman2019selectivenet}, is to let the system \emph{decline to answer}
low-confidence inputs rather than being forced to commit every time.
Early-exit and cascade systems apply a similar idea to save compute
\cite{teerapittayanon2016branchynet, lebovitz2023efficient}, but they generally assume the underlying predictor is
stable and digital. AIMC breaks that assumption: the accelerator itself
is a source of unreliable, chip-specific behavior, so the system must
decide \emph{both} which physical chip to use \emph{and} when that chip's
answer is trustworthy enough to accept.

RACE-AIMC addresses exactly this two-part problem. It does its thinking
\emph{offline} -- characterizing each accelerator, picking one winner, and
computing an exact statistical guarantee -- so that the \emph{online} path
(the part that runs at inference time, on the edge device, under a tight
energy budget) is as simple as possible: one analog inference, one
comparison, one decision (Fig.~\ref{fig:overview}).

\textbf{Contributions.} This paper makes four contributions: (i) a rule for
selecting which accelerator in a heterogeneous pool should be the
``front line'' one, based on how much useful work it does per unit of
energy; (ii) a cheap, per-accelerator stopping rule for deciding when to
trust an answer; (iii) an \emph{exact} statistical certificate on the
error rate among accepted answers, computed on data that was never used
for tuning; and (iv) transparent simulation experiments confirming the approach works.

\section{Background and Related Work}

\subsection{Computations in crossbar arrays}

A crossbar array is a grid of memory devices, one at every intersection of
a set of row and column wires. Each device's electrical conductance
encodes one neural-network weight. Applying an input voltage on each row
and reading the resulting current on each column computes, in one physical
step, a full matrix-vector product -- the workhorse operation of a neural
network layer. The quality of that computed product depends on many
pieces of the pipeline working together: how weights are encoded into
conductances, how precisely they are written (\emph{programmed}), how
noisy the read-out is, the resolution of the analog-to-digital and
digital-to-analog converters, and any circuit-level non-idealities in the
periphery of the array \cite{sebastian2020memory, le202364, rasch2023hardware, wang2022device, yousuf2025robust}.

A large body of prior work makes an analog-mapped network \emph{more
accurate for a fixed, already-built chip} -- for instance,
\emph{hardware-aware training} that trains the network to already tolerate
a chip's known noise, or \emph{fault-tolerant ensembles} that spread a
computation redundantly across devices \cite{rasch2023hardware,yousuf2025layer}. RACE-AIMC asks a
different, complementary question: given a chip that has already been
characterized, \emph{should it be trusted with this particular input right
now}, or should the query be routed elsewhere? This work uses a simulation tool, XBTorch
\cite{yousuf2026xbtorch}, because it lets ordinary PyTorch code run unchanged outside a
clearly-marked ``analog boundary,'' while still modeling persistent,
stateful conductance arrays (i.e., the same simulated chip keeps its own
noise pattern across many calls, rather than being re-randomized every
time -- which matters because real chips do not re-randomize themselves
between inferences).

A transparent multilevel weight encoding is used, rather than
the cruder binary/sign-only encoding sometimes used in the literature,
specifically so that the experiments measure the effect of the proposed policy and not an artifact of using limited bits per
weight.

\subsection{Selective classification and conformal prediction}

{Selective classification} \cite{geifman2019selectivenet} formalizes the idea
of a classifier that may abstain: instead of forcing a prediction on every
input, it accepts inputs above some confidence level and rejects the
rest, trading \emph{coverage} (how often it answers) against \emph{risk}
(how often it is wrong when it does answer). {Early-exit} and
{cascade} systems \cite{teerapittayanon2016branchynet,lebovitz2023efficient}, and heterogeneous-hardware
schedulers \cite{boroumand2021mitigating,dagli2021multi}, apply a related idea to save computation,
but typically assume the underlying model's confidence signal is stable
and digital -- an assumption that does not hold for a noisy analog chip.
{Conformal prediction} \cite{angelopoulos2023conformal, angelopoulos2024conformal} is a more general
family of tools for wrapping any predictor with a distribution-free
statistical guarantee.

RACE-AIMC's statistical target is narrower and simpler than general
conformal risk control: once a single accelerator and threshold have been
frozen, the only remaining question is a plain {Bernoulli
error rate} -- of the accepted predictions from \emph{this one fixed
subset}, what fraction are wrong? Because that target is so specific, an
\emph{exact} one-sided binomial confidence interval (rather than an
approximate or asymptotic one) can certify it directly, whereas
general-purpose conformal methods are built to control a broader,
joint notion of miscoverage.

\section{RACE-AIMC}

\subsection{System model and data separation}

Let $A = \{1,\dots,M\}$ denote a pool of $M$ AIMC
accelerators (potentially heterogeneous), all storing the same trained classifier. Accelerator $i$
has an activation energy $E_i$, a communication cost $C_i$, and, for an
input $x$, produces logits $z_i(x)\in\mathbb{R}^K$ (the raw, pre-softmax
class scores, one per class). A shared, purely digital ``front end''
computes the input features that all accelerators share. The predicted
label is the class with the highest logit,
$\hat y_i(x)=\arg\max_k z_{i,k}(x)$, and $s_i(x)$ denotes a confidence
score (defined later) for that prediction.

\paragraph{The accept/reject rule} Every accelerator has its own
confidence threshold $\tau_i$. An answer is accepted only if the
confidence score clears that threshold:
\begin{equation}
g_i(x;\tau_i) = \mathbf{1}\{s_i(x) \ge \tau_i\}.
\label{eq:accept}
\end{equation}
Here $\mathbf{1}\{\cdot\}$ is the \emph{indicator function}: it equals $1$
if the statement inside the braces is true, and $0$ otherwise. So
$g_i(x;\tau_i)$ is simply a yes/no flag -- ``accept'' or ``defer'' -- for
one input on one accelerator.

Two quantities describe how well a given threshold behaves:
\begin{align}
\phi_i(\tau) &= P[\,g_i(X;\tau)=1\,], \label{eq:coverage}\\
R_i(\tau) &= P[\,\hat y_i(X)\neq Y \mid g_i(X;\tau)=1\,]. \label{eq:risk}
\end{align}

Equation~\eqref{eq:coverage} is the {coverage}: the fraction of all
possible inputs that the accelerator is willing to answer (as opposed to
deferring). Equation~\eqref{eq:risk} is the {selective risk}: of
\emph{only the inputs it chose to answer}, what fraction are wrong? Note
this is a conditional probability -- it completely ignores the inputs that
were deferred.

The goal is to keep coverage as high as possible (answer as many inputs
directly as we can, to save energy on the fallback path) while keeping
selective risk below an application-defined budget $r$ (e.g., $r=10\%$).

\paragraph{Three independent roles for calibration data} To make the
final risk guarantee trustworthy, calibration data is split into three
\emph{non-overlapping} groups, each used for exactly one purpose:
\begin{itemize}
\item \textbf{Hardware split $H$}: estimates a persistent, per-accelerator
correction (Sec.~\ref{sec:hwcorrection}). Never used to tune thresholds
or pick an accelerator.
\item \textbf{Policy split $P$}: fits each accelerator's threshold and
selects which single accelerator to deploy (Sec.~\ref{sec:threshold}).
\item \textbf{Certificate split $C$}: evaluates the already-frozen policy
\emph{exactly once}, producing the final statistical guarantee. 
\end{itemize}

\subsection{Multilevel differential weight mapping}
\label{sec:mapping}

Analog devices can only hold a limited number of distinct conductance
levels -- here, 256 levels (8-bit resolution). Each software weight must
therefore be rounded onto this discrete grid. For a layer $\ell$, let
$\gamma_\ell = \max_{jk}|W_{\ell,jk}|$ be the largest-magnitude weight in
that layer, and let $Q = 2^8-1 = 255$. Each weight magnitude is
normalized and quantized as
\begin{equation}
q(w) = \frac{\mathrm{round}(Q\,|w|/\gamma_\ell)}{Q}.
\label{eq:quantize}
\end{equation}

Because analog conductances cannot be negative, sign is encoded using a
\emph{differential pair} of two physical devices per weight. For a
positive weight, the positive-branch conductance is
$G_{\min} + q(w)(G_{\max}-G_{\min})$ and the negative-branch conductance
sits at its floor value $G_{\min}$; the roles swap for a negative weight.
The weight is then reconstructed from the pair as
\begin{equation}
\hat w = \gamma_\ell\,\frac{G^{+}-G^{-}}{G_{\max}-G_{\min}}.
\label{eq:reconstruct}
\end{equation}

The very same per-layer scale $\gamma_\ell$ is also applied to that
layer's input during the physical matrix-vector multiply, which avoids
accidentally applying two different, mismatched scale factors (one for
weights, one for inputs) to the same computation. Every mapped layer is
audited for the number of distinct reconstructed values actually used,
any clipping, and standard mapping-error metrics (MAE, RMSE, relative
$\ell_2$ error).

\subsection{Hardware correction and confidence}
\label{sec:hwcorrection}

Every physical accelerator tends to have a persistent, systematic bias in
its output logits -- a consistent skew caused by its own programming
imperfections, distinct from random noise. Using the hardware split $H$,
this per-class bias is estimated by comparing each accelerator's raw
logits $z_i^{\mathrm{raw}}(x)$ against the reference (clean, digital)
logits $z_0(x)$:
\begin{equation}
b_i = |H|^{-1}\!\!\sum_{x\in H}\!\big(z_i^{\mathrm{raw}}(x)-z_0(x)\big),
\quad z_i(x) = z_i^{\mathrm{raw}}(x)-b_i.
\label{eq:bias}
\end{equation}

RACE-AIMC uses the {top-two logit margin} as its confidence score:
\begin{equation}
s_i(x) = z_{i,(1)}(x) - z_{i,(2)}(x),
\label{eq:margin}
\end{equation}
where $z_{i,(1)}\ge z_{i,(2)}$ are the largest and second-largest
(bias-corrected) logits for input $x$.

\subsection{Threshold fitting, selection, and certification}
\label{sec:threshold}

\paragraph{Fitting a threshold} For each accelerator $i$, policy-split
examples are sorted by decreasing confidence score, and acceptance is
built up as a growing \emph{prefix} of the most-confident examples. If a
candidate prefix accepts $n$ examples and $e$ of them are wrong, the
{Clopper-Pearson upper bound} on the true error rate is
\begin{equation}
U(e,n;\delta) = F^{-1}_{\mathrm{Beta}(e+1,\,n-e)}(1-\delta),
\label{eq:clopper}
\end{equation}
with $U=1$ by convention when $n=0$ or $e=n$.

The largest prefix satisfying $U(e,n;\delta_d)\le r_d$ is kept, where
$r_d$ is a stricter \emph{design} risk target than the true deployment
target $r$ (e.g., $r_d=8\%$ vs.\ $r=10\%$). This built-in safety margin
matters because the threshold is fit on the policy split, while the
\emph{real} certificate will later be computed fresh on the untouched
certificate split -- fitting with some slack makes it much more likely
that the final, independently-computed certificate still clears the true
target.

\paragraph{Selecting one accelerator} Once every accelerator has its own
fitted threshold and measured (empirical) coverage $\hat\phi_i$, a single
``front-line'' accelerator is chosen by a coverage-per-energy utility:
\begin{equation}
u_i = \frac{\hat\phi_i}{E_i+C_i}, \qquad i^{*} = \arg\max_i u_i.
\label{eq:utility}
\end{equation}

The winning accelerator $i^{*}$ and
its threshold $\tau_{i^{*}}$ are then \emph{frozen} -- no further tuning.

\paragraph{Certification} The frozen policy is evaluated exactly once on
the untouched certificate split $C$. If it accepts $n_C>0$ examples with
$e_C$ errors, and $U(e_C,n_C;\delta_C)\le r$, the policy is
{certified}: assuming the certificate and deployment data are
exchangeable (drawn from the same underlying distribution), this is a
rigorous, finite-sample confidence statement about the true, real-world
selective risk of the frozen policy -- not merely an estimate.

\section{Experimental Methodology}

\subsection{Model, data, and the analog/digital boundary}

Experiments use CIFAR-10. The feature-extraction path is a small
convolutional network: Conv(3,32,3)-ReLU-MaxPool,
Conv(32,64,3)-ReLU-MaxPool, Conv(64,128,3)-ReLU-adaptive-pool-to-$4\times
4$, followed by a classifier head Linear(2048,256)-ReLU-Linear(256,10).
Each of the (multiple, independently seeded) trained realizations uses
40{,}000 training examples for 30 epochs, batch size 128, AdamW
(learning rate $2\times10^{-3}$, weight decay $5\times10^{-4}$), cosine
annealing, cross-entropy loss, and standard random-crop/flip
augmentation. The remaining 10{,}000 training examples are split into
$|H|=3000$, $|P|=2000$, $|C|=5000$; the official 10{,}000-example test
set is used only descriptively (i.e., to sanity-check behavior, not to
compute the certified guarantee).

\subsection{Heterogeneous accelerator population}

Experiments use PyTorch 2.7.0, torchvision 0.22.0, XBTorch 1.0.0, and an
NVIDIA A100 GPU (for simulation). All six simulated accelerator profiles
share 8-bit conductance devices, 8-bit ADC/DAC converters,
$G_{\min}=133$ $\mu S$, $G_{\max}=233$ $\mu S$, a $0.3\,$V read voltage, instantaneous
input encoding, and a $2500\times2500$ physical
device array. Table~\ref{tab:profiles} lists how noise, fault rate, and
energy/latency vary across the six profiles (A0, the mildest and
cheapest, through A5, the harshest and most expensive). Each profile uses
its own deterministic-but-distinct simulator seed, so the six profiles
form a broad, reproducible \emph{stress envelope} for testing the
scheduling method, rather than a fit to any one specific real chip.

\begin{table}[htbp]
\centering
\caption{Heterogeneous 8-bit AIMC accelerator profiles}
\label{tab:profiles}
\resizebox{\columnwidth}{!}{%
\begin{tabular}{lccccc}
\toprule
Profile & \begin{tabular}[c]{@{}c@{}}Read noise\\ ($\mu$S)\end{tabular} & \begin{tabular}[c]{@{}c@{}}Write noise\\ ($\mu$J)\end{tabular} & \begin{tabular}[c]{@{}c@{}}Stuck\\ (\%)\end{tabular} & \begin{tabular}[c]{@{}c@{}}$E$\\ ($\mu$J)\end{tabular} & \begin{tabular}[c]{@{}c@{}}$L$\\ (ms)\end{tabular} \\
\midrule
A0 & 1.0  & 2.0  & 0.020 & 0.7 & 0.090 \\
A1 & 2.8  & 7.6  & 0.116 & 0.8 & 0.105 \\
A2 & 4.6  & 13.2 & 0.212 & 0.9 & 0.120 \\
A3 & 6.4  & 18.8 & 0.308 & 1.0 & 0.135 \\
A4 & 8.2  & 24.4 & 0.404 & 1.1 & 0.150 \\
A5 & 10.0 & 30.0 & 0.500 & 1.2 & 0.165 \\
\bottomrule
\end{tabular}%
}
\end{table}

\subsection{Frozen protocol, baselines, and energy model}

The deployment risk target is $r=0.10$, with a stricter design target
$r_d=0.08$, $\delta_d=0.10$, and $\delta_C=0.05$. Five confirmatory
random seeds were fixed \emph{before} any of
these runs were executed, so the reported results are genuine
out-of-sample confirmations rather than results selected after the fact.

Four systems are compared: (1) clean digital inference (the ideal,
no-analog-error baseline), (2) forced execution of only the selected
analog accelerator (no deferral), (3) the full selective system with a
digital fallback for deferred queries, and (4) a uniform always-on
ensemble that activates \emph{all six} analog accelerators and averages
their outputs. The energy model includes analog classifier activation
energy, a $0.12\,\mu$J communication cost per accelerator, a
$0.003\,\mu$J controller cost, $0.0002\,\mu$J per fused class, and a
modeled $4\,\mu$J fallback cost; the (shared, digital) convolutional path
is excluded from this comparison since it is identical across all four
systems. This makes the energy comparison a component-level modeled
estimate, not a full-platform hardware measurement.

\section{Results}

% \subsection{Is the weight mapping trustworthy?}

% Before trusting any risk certificate, it is worth checking that the
% 8-bit weight mapping itself is behaving sensibly (i.e., that later
% results reflect the proposed \emph{scheduling policy}, not a broken or
% overly crude weight encoding). Across the five confirmatory runs, the
% large linear layer used 358--407 distinct signed reconstructed weight
% values with zero clipping, a mapping mean-absolute-error of
% $7.59\times10^{-4}$, and a relative $\ell_2$ mapping error of
% $1.18\%\pm0.15\%$; the output layer showed similarly small errors. On a
% development check, an \emph{ideal} (noise-free) multilevel-mapped
% accelerator lost only 0.20 accuracy points versus clean digital
% inference; adding the selected accelerator's real hardware noise cost a
% further 0.26 points. In other words, the mapping itself is not the
% source of the interesting behavior in this paper -- the accelerators'
% physical nonidealities are.

\subsection{Selection beats blind redundancy}

Nonidealities remain meaningful across the pool: on the confirmatory
runs, the selected accelerator alone reaches $81.00\%\pm0.55\%$ forced
accuracy versus $81.40\%\pm0.50\%$ for clean digital inference. In
contrast, blindly averaging \emph{all six} heterogeneous accelerators
falls to only $75.32\%\pm1.74\%$, because the harshest profiles drag the
combined answer down with poor-quality logits. Selecting a single
good accelerator therefore outperforms blind, always-on redundancy by
5.68 accuracy points on average -- despite using far less energy (see
Sec.~\ref{sec:energy}).

\subsection{Independent risk certificates and accuracy}

\begin{table}[htbp]
\centering
\caption{Summary of simulation results (percentages).}
\label{tab:confirmatory}
\resizebox{\columnwidth}{!}{%
\begin{tabular}{lccccc}
\toprule
Run & \begin{tabular}[c]{@{}c@{}}Cert.\\ cov.\end{tabular} & \begin{tabular}[c]{@{}c@{}}Cert.\\ risk\end{tabular} & \begin{tabular}[c]{@{}c@{}}95\%\\ upper\end{tabular} & \begin{tabular}[c]{@{}c@{}}Test\\ cov.\end{tabular} & \begin{tabular}[c]{@{}c@{}}Hybrid\\ acc.\end{tabular} \\
\midrule
R1  & 70.90 & 5.73 & 6.41 & 70.46 & 82.06 \\
R2  & 69.44 & 7.14 & 7.90 & 69.35 & 80.68 \\
R3  & 72.20 & 8.09 & 8.87 & 71.70 & 81.42 \\
R4  & 70.88 & 7.20 & 7.95 & 70.98 & 81.57 \\
R5  & 70.96 & 7.24 & 8.00 & 70.83 & 81.27 \\
\midrule
\textbf{Mean}  & \textbf{70.88} & \textbf{7.08} & \textbf{7.83} & \textbf{70.66} & \textbf{81.40} \\
\bottomrule
\end{tabular}%
}
\end{table}

\begin{figure*}[t]
\centering
\includegraphics[width=0.8\textwidth]{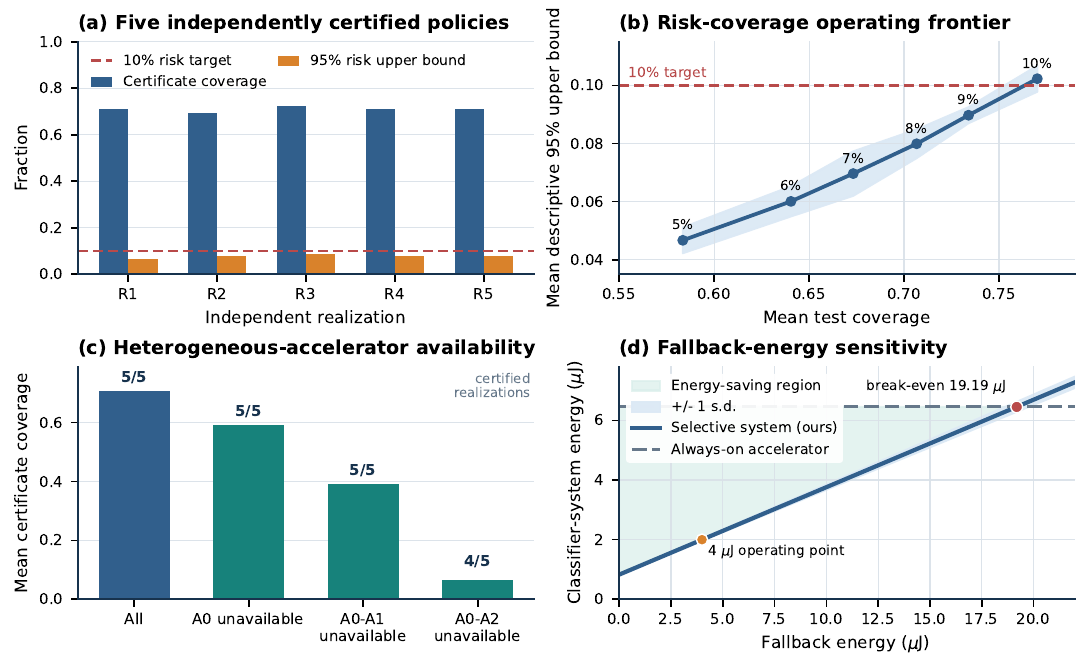}
\caption{RACE-AIMC simulation results. (a) Independent frozen realizations;
all one-sided certificate bounds are below the 10\% target. (b) Post-hoc
risk-coverage frontier; labels are design-risk settings and shading is one
standard deviation. (c) Reconfiguration as low-severity accelerator profiles
become unavailable; labels give independently certified realizations. (d)
Modeled energy versus fallback cost, including across-run variation, the
reported operating point, and the energy-saving region.}
\label{fig:results}
\end{figure*}

All five frozen policies certify below the 10\% risk target
(Table~\ref{tab:confirmatory}, Fig.~\ref{fig:results}(a)). The mean
certified coverage is $70.88\%\pm0.98\%$ (i.e., about 71\% of test
queries are answered directly by the analog accelerator), the mean
accepted empirical risk is $7.08\%\pm0.85\%$, and the mean one-sided 95\%
upper bound on that risk is $7.83\%\pm0.89\%$ (worst case across the five
runs: 8.87\%). It is important to be precise about what this means
statistically: these are five \emph{individual} 95\% certificates, not
one pooled or simultaneous guarantee across all five runs together.
Descriptive results on the separate, held-out test set (70.66\% coverage,
7.46\% accepted error) closely track the certified numbers, which is a
good sign that the certificate generalizes.

Rejected (deferred) queries fall back to the digital path. The resulting
{hybrid} system reaches $81.40\%\pm0.50\%$ overall accuracy --
statistically indistinguishable from clean digital inference
(mean difference: 0.004 points) -- while accepted-only test accuracy on
the confident, analog-answered subset alone reaches 92.54\%.

\subsection{Risk, availability, and confidence-score ablations}

\paragraph{Risk-coverage tradeoff.} Sweeping the design risk target shows
the expected tradeoff: at a stricter 5\% design risk, mean test coverage
is only 58.35\%, but by 9\% design risk it rises to 73.40\%, with all
five certificate replays still passing (Fig.~\ref{fig:results}(b)). Push
the design risk all the way to the 10\% deployment target itself, and
coverage climbs to 77.02\% -- but only 3 of 5 certificates pass. This
illustrates concretely why RACE-AIMC deliberately fits thresholds at a
\emph{stricter} design risk than the deployment target: without that
safety margin, the certificate becomes unreliable more often.

\paragraph{Accelerator availability.} With the full pool available, the
mildest, cheapest accelerator (A0) is chosen in every single run, since
it dominates on both accuracy and cost. To test robustness, A0 (then
A0--A1, then A0--A2) were made artificially unavailable, forcing the
system to reselect from the remaining, harsher accelerators
(Fig.~\ref{fig:results}(c)). Mean certified coverage drops accordingly --
to 59.27\%, 39.24\%, and 6.47\% -- with 5/5, 5/5, and 4/5 certificates
still passing, respectively. The system degrades gracefully under
moderate hardware loss, but the harshest accelerators alone offer only
limited useful coverage.

\paragraph{Confidence-score choice.} Table~\ref{tab:scores} replays the
same protocol using four alternative confidence scores in place of the
primary logit margin: maximum softmax probability, probability margin,
log-odds, and negative entropy. All four alternatives still certify 5/5
policies, with mean test coverage in a narrow 70.53--71.69\% band and
mean accepted risk in a narrow 7.29--7.53\% band. This supports the
finding that the method is robust to the specific choice of confidence
score.

\begin{table}[t]
\centering
\caption{Post-hoc confidence-score replay (mean \%, five runs).}
\label{tab:scores}
\small
\setlength{\tabcolsep}{2.8pt}
\begin{tabular}{lcccc}
\toprule
Score & Cert.\ cov. & 95\% upper & Test cov. & Test risk\\
\midrule
Max softmax          & 71.70 & 7.82 & 71.69 & 7.45\\
Probability margin    & 71.69 & 7.90 & 71.40 & 7.53\\
Logit margin (primary)& 70.88 & 7.83 & 70.66 & 7.46\\
Log-odds              & 71.70 & 7.82 & 71.69 & 7.45\\
Negative entropy       & 70.76 & 7.87 & 70.53 & 7.29\\
\bottomrule
\end{tabular}
\end{table}

\subsection{Energy sensitivity}
\label{sec:energy}

At the modeled $4\,\mu$J fallback cost, the full selective system
consumes $2.00\pm0.03\,\mu$J per query, versus $6.45\,\mu$J for the
always-on, six-accelerator baseline -- a modeled reduction of
$69.02\%\pm0.53\%$. Ignoring fallback cost entirely, a single selected
activation alone costs just 0.825 $\mu$J, 87.21\% below the always-on
baseline. Since the true cost of the digital fallback path is inherently
platform-dependent, Fig.~\ref{fig:results}(d) sweeps this cost directly:
the selective system remains cheaper than the always-on baseline right up
until fallback cost reaches a mean break-even point of
$19.19\pm0.56\,\mu$J -- well above the $4\,\mu$J modeled operating point
used elsewhere in this paper.

\section{Discussion}

RACE-AIMC treats analog hardware reliability as something to be
\emph{managed with a certified decision}, rather than a property every
accelerator must be forced to imitate perfectly. The entire online
workload is deliberately tiny: one analog inference, a top-two logit
comparison, a subtraction, a threshold check, and a routing decision.
All the harder work -- characterizing hardware, selecting an accelerator,
and computing the statistical certificate -- happens offline, once, ahead
of deployment.

Although this paper studies a \emph{heterogeneous} pool of six
differently-flawed accelerators, the same logic applies to a
\emph{homogeneous} pool of nominally identical chips: even ``the same''
accelerator design acquires distinct programmed states, aging patterns,
faults, and operating temperatures across individual physical units.
Deliberate heterogeneity simply makes the selection opportunity larger --
it is not a requirement of the method.

Three limitations are worth being explicit about. First, the six accelerator profiles are an intentionally
broad, simulator-generated stress envelope, not statistics fitted to
measured, fabricated silicon; likewise, all energy numbers are modeled
component-level costs rather than full-chip hardware measurements. Second,
the statistical certificate relies on an \emph{exchangeability}
assumption between the certificate data and future deployment data --
it must be re-earned (i.e., recomputed) after any data distribution
shift, temperature change, conductance drift, or change to the frozen
policy itself. Third, CIFAR-10 with a compact CNN is a controlled proof
of concept, useful for isolating and testing the proposed method, but not
yet evidence of production-scale behavior.

Future work includes validating the approach on measured traces from
fabricated accelerators (rather than simulated ones), studying
drift-aware recalibration and sequential (anytime-valid) risk control, and exploring multi-stage fallback
chains together with joint latency-energy objectives.

\section{Conclusion}

Across five independent runs, RACE-AIMC's
frozen policies all certified an accepted-error rate below a 10\% target,
while directly answering roughly 71\% of queries using a single, cheap
analog accelerator, matching clean digital accuracy overall, and cutting
modeled classifier-system energy by 69\% relative to a naive always-on
heterogeneous ensemble. The central idea -- characterize hardware once,
freeze a simple decision rule, and certify it rigorously on untouched
data -- lets a lightweight edge controller expose a statistically
trustworthy region of low-cost analog operation, while reserving the more
expensive, reliable fallback path only for the queries that actually need
it.

\printbibliography

\end{document}